\documentclass[aip,cha,numerical,reprint,eqsecnum,floats]{revtex4-1}

\usepackage[english]{babel}
\usepackage{amsmath}
\usepackage{amssymb}
\usepackage{mathrsfs}
\usepackage{amsthm}
\usepackage{textcomp}
\usepackage{xcolor}
\usepackage{graphicx}
\usepackage{enumerate}
\usepackage{mathtools}
\usepackage{hyperref}
\usepackage[title]{appendix}
\usepackage{subfigure}
\usepackage{bm}

\newtheorem*{remark}{Remark}

\begin{document}

\title{The Size of the Sync Basin Revisited}

\author{Robin Delabays}
\affiliation{School of Engineering, University of Applied Sciences of Western Switzerland, CH-1950 Sion, Switzerland}
\affiliation{Section de Math\'ematiques, Universit\'e de Gen\`eve, CH-1211 Gen\`eve, Switzerland}
\author{Melvyn Tyloo}
\affiliation{School of Engineering, University of Applied Sciences of Western Switzerland, CH-1950 Sion, Switzerland}
\affiliation{Institute of Physics, \'Ecole Polytechnique F\'ed\'erale de Lausanne (EPFL), CH-1015 Lausanne, Switzerland}
\author{Philippe Jacquod}
\affiliation{School of Engineering, University of Applied Sciences of Western Switzerland, CH-1950 Sion, Switzerland}

\date{\today}

\begin{abstract}
 In dynamical systems, the full stability of fixed point solutions is determined by their basin of attraction. 
 Characterizing the structure of these basins is, in general, a complicated task, especially in high dimensionality. 
 Recent works have advocated to quantify the non-linear stability of fixed points of dynamical systems through the relative volumes of 
 the associated basins of attraction [D. A. Wiley {\it et al.} Chaos {\bf 16}, 015103 (2006), P. J. Menck {\it et al.} Nat. Phys. {\bf 9}, 89 (2013)].  
 Here we revisit this issue and propose an efficient numerical method to estimate these volumes. 
 The algorithm first identifies stable fixed points.
 Second, a set of initial conditions is considered that are randomly distributed at the surface of hypercubes centered on each fixed point. 
 These initial conditions are dynamically evolved. 
 The linear size of each basin of attraction is finally determined by the proportion of initial conditions which converge back to the fixed point. 
 Armed with this algorithm, we revisit the problem considered by Wiley et al. in a seminal paper [D. A. Wiley {\it et al.} Chaos {\bf 16}, 015103 (2006)] that inspired the title of the 
 present manuscript, and consider the equal-frequency Kuramoto model on a cycle. 
 Fixed points of this model are characterized by an integer winding number $q$ and the number $n$ of oscillators. 
 We find that the basin volumes scale as $(1-4q/n)^n$, contrasting with the Gaussian behavior postulated in Wiley et al.'s paper. 
 Finally, we show the applicability of our method to complex models of coupled oscillators with different natural frequencies and on meshed networks. 
\end{abstract}

\maketitle

\begin{quotation}
 Many natural systems of coupled elements, such as fireflies, pacemaker cells or electrical grids, exhibit synchronization phenomena. 
 When a system synchronizes, each of its components behaves coherently with respect to the others, due to the coupling between them. 
 A central issue in various fields of science and engineering is to understand how robust is this synchronized state against external perturbations or imperfections in the system. 
 Mathematically, this problem is usually hard. 
 Here we propose a tractable numerical approach which first identifies the synchronization states and second evaluates the magnitude of the largest perturbation such that the system converges 
 back to its initial synchronized state. 
 Our method allows to tackle large complex systems in a reasonable computation time with good resolution, which was not the case for numerical methods proposed so far.
\end{quotation}

\section{Introduction}
Models of coupled dynamical systems are widely used to investigate collective behaviors in complex systems. 
One particularly puzzling phenomena is that of synchrony, 
where different individual dynamical systems start to behave coherently when sufficiently strongly coupled.~\cite{Ott02,Str04,Str00,Ace05,Dor14} 
The Kuramoto model was introduced~\cite{Kur75,Kur84} to describe such synchronizing behaviors. 
The model considers a set of $n$ coupled harmonic oscillators, with angle coordinate $\theta_i$ and natural frequency $P_i$, 
\begin{align}\label{eq:kuramoto_gen}
 \dot{\theta}_i &= P_i - \sum_{j}K_{ij}\sin(\theta_i-\theta_j)\, , & i=1,...,n\, ,
\end{align}
where $K_{ij}\in\mathbb{R}$ is the coupling constant between oscillators $i$ and $j$. 
In its original formulation, the Kuramoto model considers identical all-to-all coupling, $K_{ij}\equiv K/n$.~\cite{Kur75,Kur84} 
It was found that for a coupling constant $K$ exceeding a critical value $K_c$, a finite non-empty set $F\subset\{1,...,n\}$ of oscillators synchronizes, 
i.e. $\dot{\theta}_i-\dot{\theta}_j=0$, for $i,j\in F$. 
This type of synchrony, where oscillators rotate at the same frequency, $\dot{\theta}_i=\dot{\theta}_j$, for $i,j\in F$, but not necessarily the same phase, is called 
\emph{frequency synchronization}. 
\emph{Phase synchronization}, where additionally $\theta_i=\theta_j$, $\forall i,j$, is in general not achievable for heterogeneous natural frequencies. 
In this manuscript, ``synchronization'' refers to ``frequency synchronization''.

The value of $K_c$ can be computed by solving an implicit equation.~\cite{Dor14,Aey04,Mir05,Ver08}
The critical coupling $K_c$ depends on the distribution of the natural frequencies $g(P)$. 
In particular, if the support of the distribution $g(P)$ is compact, full-synchrony, i.e. $\dot{\theta}_i-\dot{\theta}_j=0$ for all $i,j$, is reached for large enough $K$.~\cite{Erm85,Hem93}

The Kuramoto model in its various versions has evolved into a paradigm for investigating synchronizing behaviors. 
Its popularity stems from its simple formulation, which allows a tractable analytical treatment while still capturing the essence of the synchronizing behavior of many real systems, 
in fields as diverse as physics,~\cite{Wie96} chemistry,~\cite{Kur84b} biology~\cite{Som91} or electrical engineering.~\cite{Dor13}
These two advantages explain the popularity of the Kuramoto model in science and engineering, which led to many generalizations reviewed for instance in Refs.~\onlinecite{Ace05,Dor14}. 

Full frequency synchrony in the Kuramoto model is reached at a fixed point of Eq.~\eqref{eq:kuramoto_gen}. 
A natural question is then to assess the stability of these fixed points. 
Most works on the stability of fixed points of dynamical systems rely on the seminal work of Lyapunov~\cite{Lya92,Lya92b} at the end of the XIX$^{\rm th}$ century. 
Lyapunov first investigated \emph{linear stability}, by linearizing the dynamics around a given fixed point. 
For small deviations, the dynamics is determined by a stability matrix whose eigenvalues are called \emph{Lyapunov exponents}. 
The fixed point is linearly stable if all its Lyapunov exponents are non-positive. 
This guarantees that small enough deviations go exponentially fast to zero. 
Linear stability is however a local concept and gives no information about the stability of the system against large perturbations. 

Lyapunov went beyond linear stability with his second method, which assesses stability based on the existence of what is now called a \emph{Lyapunov function}.~\cite{Lya92,Lya92b} 
The latter generalizes the concept of energy for the states of a dynamical system. 
The Lyapunov function of a system can be used to determine the \emph{basin of attraction} of a given fixed point,~\cite{Pai81,Chi89} 
which is the set of all initial conditions converging dynamically to this fixed point. 
A global measure of the stability of a fixed point is given by the volume of its basin of attraction - this has been called \emph{basin stability}.~\cite{Wil06,Men13,Sch17}
Clearly, the larger the basin of attraction, the more likely it is to reach the corresponding fixed point dynamically. 
This gives a global measure of the stability of a fixed point.

\begin{figure}
 \centering
 \includegraphics[width=.4\textwidth]{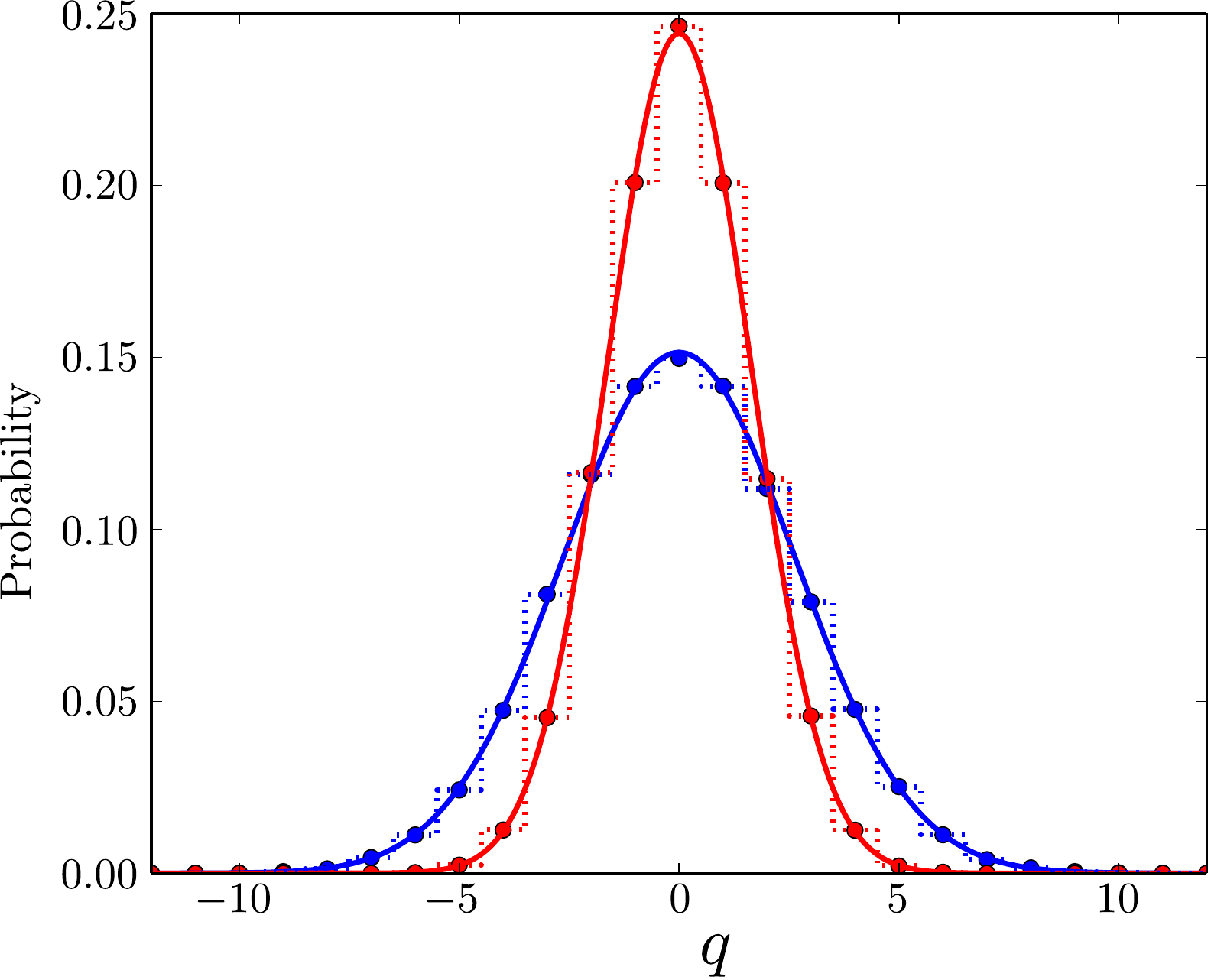}
 \caption{Distributions of the initial (blue) and final (red) winding numbers for the equal frequency Kuramoto model on a cycle [Eq.~\eqref{eq:kuramoto_cycle}] with $n=83$ nodes. 
 Initial states have been chosen randomly. Continuous curves are Gaussian fits with $\sigma=2.63$ (blue) and $\sigma=1.63$ (red).}
 \label{fig:gaussian_fit}
\end{figure}

In networks of all-to-all coupled oscillators, tight estimates of the volume of the basin of attraction of the sychronous state are known.~\cite{Dor11}
Much less is known about the basins of attraction of cycle networks.
As pointed out by Korsak~\cite{Kor72} already in 1972 in the context of electrical networks,
the Kuramoto model on a cycle network admits several stable fixed points, characterized by their winding numbers (to be defined in Sec.~\ref{sec:single_cycle}).
In Ref.~\onlinecite{Wil06}, Wiley et al. considered such network topology, with identical frequencies and 
investigated how the volume of the basin of attraction of a stable fixed point is related to its winding number. 
In particular, they are interested in the likelihood of the system to reach the phase synchronous state. 
This likelihood is directly related to the volume of the basin of attraction of the phase synchronous state, which they call ``sync basin''.
Starting from random initial conditions, they numerically evolved the system until it converged to a stable fixed point. 
The volume of the basin of attraction of every fixed point was then estimated by the proportion of initial conditions that converged to it. 
It was found that the volume of the basins of attraction follow a Gaussian distribution with respect to the winding numbers $q$, as shown in Fig.~\ref{fig:gaussian_fit} (red dots). 
One issue with that procedure is that the winding numbers of randomly chosen initial conditions also follow a Gaussian distribution, 
\begin{align}\label{eq:gaussian_fit}
p(q) = (\sqrt{2\pi}\sigma)^{-1}\exp\left(-q^2/2\sigma^2\right)\, .
\end{align}
From the data shown in Fig.~\ref{fig:gaussian_fit}, we obtain a standard deviations of $\sigma=2.63$ for the distribution of winding numbers of the initial conditions and 
a narrower distribution with $\sigma=1.63$ for the converged fixed points (the latter value in agreement with Ref.~\onlinecite{Wil06}). 
The Gaussian distribution for the initial conditions can easily be understood once one realizes that picking an initial condition is similar to a random walk.~\cite{Pat11} 
The node index along the cycle corresponds to a time step index and the angle on each node gives by how much and in what direction the random walk progresses. 
Large winding numbers correspond then to random walks with large excursions. This analogy explains the obtained Gaussian distribution for initial winding numbers.

We also observe that the winding number of the initial conditions and of the converged fixed points are significantly correlated, with a correlation coefficient of 0.47. 
Therefore, if one does not have enough resolution for the initial conditions, the distribution of the winding number of the final states may, at least partially, 
reflect the initial distribution of $q$ instead of the volume of the basins of attraction. 
Due to the high dimensionality of the state space ($n=83$ in Fig.~\ref{fig:gaussian_fit} and $n=80$ in Ref.~\onlinecite{Wil06}), 
simulations with random initial conditions would need an unfeasible number of runs to representatively cover the whole state space.

To the best of our knowledge, the only paper, beside Ref.~\onlinecite{Wil06}, focusing specifically on the basins of attraction of the Kuramoto model on cycle network is Ref.~\onlinecite{Ha12}, 
which analytically obtains lower bounds on the volume of the basins of attraction for a cycle of Kuramoto oscillators with unidirectional coupling.

We therefore revisit this issue by constructing a new systematic numerical method. 
Our approach is first to find all the stable fixed points of Eq.~\eqref{eq:kuramoto_gen} describing all the possible frequency-synchronous states, 
and second to perturb them in random directions with an increasing magnitude to assess the volume of their basins of attraction. 
The volume is estimated from the magnitude of the largest perturbation still converging to the initial fixed point. 
In the case of a single cycle with identical frequencies, one can analytically identify all stable fixed points and the problem is sufficiently tractable to obtain an analytical estimate 
of the volume of the basins of attraction, which we confirm numerically. 
We show that for $n\gg 1$ and $q$ not too small, the volume of the basins of attraction scales as $V_q\sim (1-4q/n)^n$ instead of the Gaussian law of Wiley et al. [Eq.~\eqref{eq:gaussian_fit}]. 
We then extend our perturbation procedure to cycle networks with non-identical frequencies and to meshed networks with identical frequencies. 
Our numerical method guarantees that we investigate every basin of attraction with a representative number of initial conditions.
It is based on (i) a numerical procedure to systematically find stable fixed points of Eq.~\eqref{eq:kuramoto_gen} on any meshed network and (ii) the perturbation procedure described above. 
We believe that our method could be applied to other fields of research such as planar spin glasses~\cite{Mez87} or disordered Josephson junction arrays~\cite{Faz01} in condensed matter physics.

\section{Single cycle}\label{sec:single_cycle}
We first revisit the model of Ref.~\onlinecite{Wil06} and consider the Kuramoto model on a cycle with $n$ nodes, equal frequencies $P_i\equiv P_0$, for all $i$, 
and identical coupling $K_{ij}\equiv K$, for all connected nodes $i$ and $j$. 
In a frame rotating with angular frequency $P_0$, after the change of variables $\theta_i\to\theta_i+P_0t$, Eq.~\eqref{eq:kuramoto_gen} reduces to
\begin{align}\label{eq:kuramoto_cycle}
 \dot{\theta}_i &= - K\sin(\theta_i-\theta_{i-1}) - K\sin(\theta_i-\theta_{i+1})\, , & i=1,...,n\, ,
\end{align}
where indices are taken modulo $n$.

We define the angle difference $\Delta_{ij}\coloneqq\theta_i-\theta_j$ taken modulo $2\pi$ in the interval $(-\pi,\pi]$. 
Fixed points of Eq.~\eqref{eq:kuramoto_cycle} satisfy either
\begin{align}\label{eq:angle_diff}
 \Delta_{i,i+1} &= \Delta_{i-1,i} & &\text{or}&  \Delta_{i,i+1}&=\pm\pi - \Delta_{i-1,i}\, , 
\end{align}
where the sign in front of $\pi$ in the right-hand side is chosen to ensure that $\Delta_{i,i+1}\in(-\pi,\pi]$.

Given an angle vector $\vec{\theta}=(\theta_1,...,\theta_n)$ we define the integer \emph{winding number} on the cycle 
\begin{align}\label{eq:winding}
 q(\vec{\theta}) &\coloneqq (2\pi)^{-1}\sum_{k = 1}^n\Delta_{{k+1},{k}}\, .
\end{align}
When summing the angle differences around the cycle, we have to end at an integer multiple of $2\pi$ to guarantee single-valuedness of angles. 
Therefore, $q(\vec{\theta})\in\mathbb{Z}$. 

\begin{remark}
 For any graph topology, we can always define a winding number on every cycle. 
 A fixed point is then characterized by a \emph{winding vector}, whose components are the winding numbers on each cycle.
\end{remark}

According to Ref.~\onlinecite{Del16}, any stable fixed point of Eq.~\eqref{eq:kuramoto_cycle} must have all angle differences between neighboring oscillators in $[-\pi/2,\pi/2]$. 
Eq.~\eqref{eq:angle_diff} then implies that there is a unique stable fixed point, $\vec{\theta}^{(q)}$, with winding number $q$,
\begin{align}\label{eq:sfp}
 \Delta = 2\pi q/n &\iff \theta_i^{(q)} = 2\pi q i/n + \theta_0^{(q)}\, ,
\end{align}
for all $i=1,...,n$, where $\theta_0^{(q)}$ is an arbitrary uniform angle shift. 
This implies that for a fixed point to be stable, the winding number cannot be larger than $q_{\max}\coloneqq{\rm Int}(n/4)$, which would imply angle differences larger than $\pi/2$ otherwise. 
Winding numbers for stable fixed points then range from $-q_{\max}$ to $+q_{\max}$, which gives $2q_{\max}+1$ stable fixed points,~\cite{Del16} 
with the caveat that if $n$ is a multiple of $4$, the fixed point with winding number $q_{\max}$ has all Lyapunov exponents equal to zero, and its basin of attraction has measure zero.~\cite{Man16}

A fixed point of Eq.~\eqref{eq:kuramoto_cycle} is unstable if one~\cite{Del16} or more~\cite{Tay12} angle differences are larger than $\pi/2$. 
In this case, a fixed point has $n-j$ angle differences $\Delta_{k+1,k}\equiv\Delta\in[-\pi/2,\pi/2]$ and $j$ angle differences $\Delta_{k+1,k}\equiv\pm\pi-\Delta$, 
with $j>0$, and $\sum_k\Delta_{k+1,k}=2\pi q$.

\subsection{Identical frequencies: Analytical approach}\label{sec:analytic}
For a cycle network of length $n$ with identical frequencies, we derive an analytical expression for the volume of the basins of attraction. 
Our approach is to approximate the basin of attraction of a given stable fixed point by the hypercube centered at the fixed point 
and whose radius is the distance to the closest unstable fixed point. 

As Eq.~\eqref{eq:kuramoto_cycle} is invariant under a constant shift of all angles $\theta_0^{(q)}$, we will work in the hyperplane ${\cal H}_{n-1}$ orthogonal to the vector $(1,...,1)$. 
The angle vector $\vec{\theta}^{(q)}$ of Eq.~\eqref{eq:sfp} projected on ${\cal H}_{n-1}$ has components
\begin{align}\label{eq:angle_vec}
 \theta_i^{(q)} &= \frac{2\pi q}{n} i - \frac{n-1}{n}q\pi\, .
\end{align}

According to Ref.~\onlinecite{Dev12}, a fixed point of Eq.~\eqref{eq:kuramoto_cycle} has a unique unstable direction in angle space if and only if 
it has a single angle difference between neighboring oscillators which is larger than $\pi/2$. 
Such an unstable fixed point is called a {\it 1-saddle} point. 
Consider then a 1-saddle point with winding number $q'$, where the $k^{\text{th}}$ angle difference is larger than $\pi/2$. 
Combining Eqs.~\eqref{eq:angle_diff} and \eqref{eq:winding}, its winding number $q'$ is given by 
\begin{align}
 (n-1)\cdot\Delta'+\pi-\Delta' = 2\pi q' &\iff \Delta'=\frac{2q'-1}{n-2}\pi\, .
\end{align}
This allows to compute the components of the 1-saddle angle vector, $\vec{\varphi}^{(q')}$, projected on ${\cal H}_{n-1}$ (see Appendix~\ref{ap:1saddle}),
\begin{align}
 \varphi_i^{(q')} &= \pi\left[\frac{2q'-1}{n-2}i+\frac{-2n^2k+2nk-8q'k-n}{2n(n-2)} + T_{i}^{(k)}\right]\, ,
\end{align}
where
\begin{align}
 T_{i}^{(k)} &= \left\{
  \begin{array}{ll}
   \frac{10nq'-n^2}{2n(n-2)}\, , & \text{if }i<k\, ,\\
   \frac{2nq'+n^2}{2n(n-2)}\, , & \text{if }i\geq k\, .
  \end{array}
  \right.
\end{align}

We have found numerically that stable fixed points and 1-saddles are closest when they have the same winding number (see Appendix~\ref{ap:fixed_point_dist}). 
We thus investigate the case $q=q'$. 
The difference between angles is easily obtained as
\begin{align}
 \theta_i^{(q)}-\varphi_i^{(q)} &= \left\{
 \begin{array}{ll}
  \frac{(1+2i-2k+n)(n-4q)}{2(n-2)n}\pi\, , & \text{if }i<k\, , \\
  \frac{(1+2i-2k-n)(n-4q)}{2(n-2)n}\pi\, , & \text{if }i\geq k\, ,
 \end{array}
 \right. \label{eq:d_theta_i}
\end{align}
which gives the distance between the stable fixed point and the unstable 1-saddle point
\begin{align}\label{eq:distance}
 \|\vec{\theta}^{(q)}-\vec{\varphi}^{(q)}\|_{\infty} &= \frac{(n-1)(n-4q)}{2(n-2)n}\pi\, .
\end{align}
In particular, the large $n$ limit is 
\begin{align}
 \lim_{n\to\infty}\|\vec{\theta}^{(q)}-\vec{\varphi}^{(q)}\|_{\infty} &= \frac{\pi}{2}\left(1-\frac{q}{q_{\max}}\right)\, .
\end{align}
We remark that we lost the dependence on $k$, meaning that the stable fixed point $\vec{\theta}^{(q)}$ is equidistant to all 1-saddles with the same winding number. 
This indicates that the basins of attraction are isotropic in the directions of the 1-saddle points. 

Since $\vec{\theta}^{(q)}$ is equidistant to all 1-saddles, one expect, for $n\gg1$, that the volume $V_q$ of the basin of attraction is well approximated by an hypercube of side 
$\|\vec{\theta}^{(q)}-\vec{\varphi}^{(q)}\|_\infty$, up to a constant factor, i.e. 
\begin{align}\label{eq:scaling}
 V_q &\sim\|\vec{\theta}^{(q)}-\vec{\varphi}^{(q)}\|_\infty^n\sim(1-q/q_{\max})^n\, .
\end{align}
As $q_{\max}={\rm Int}(n/4)$, $V_q\to e^{-4q}$ in the limit $n\to\infty$ for fixed $q$. 

We further found numerically that 1-saddles are closer to the stable fixed points than $p$-saddles, $p>1$ (see Appendix~\ref{ap:fixed_point_dist}). 
This suggests that Eq.~\eqref{eq:scaling} underestimates the volume of basins of attraction. 

\begin{remark}
 In our convention, angle differences are taken in the interval $(-\pi,\pi]$. 
 Trying to construct a 1-saddle with $q'=0$, one obtains
 \begin{align}
  \Delta' = -\frac{\pi}{n-2} &\iff \pi-\Delta' = \frac{n-1}{n-2}\pi > \pi\, .
 \end{align}
 The angle vector obtained in this way has winding number $q'=-1$. 
 This means that there is no 1-saddle with winding number zero.
 Eq.~\eqref{eq:d_theta_i} then applies to $q>0$.
 It can be checked that there are no $p$-saddles with winding number $q=0$ for any $p\geq 1$. 
\end{remark}

\subsection{Identical frequencies: Numerical approach}\label{sec:numeric_cyc_0}
To validate the scaling of Eq.~\eqref{eq:scaling}, we numerically estimate the volume of the basin of attraction of each stable fixed point of Eq.~\eqref{eq:kuramoto_cycle}. 
For a cycle network of length $n$ with identical frequencies, all stable fixed points are known and given by Eq.~\eqref{eq:angle_vec}. 
To estimate the volume of the basin of attraction of each $\vec{\theta}^{(q)}$, we randomly choose $d$ normalized perturbation vectors $\vec{\epsilon}_j\in{\cal H}_{n-1}\subset\mathbb{R}^n$, 
for $j=1,...,d$, $\|\vec{\epsilon}_j\|_{\infty}=1$. 
We then consider perturbed states 
\begin{align}\label{eq:perturbed_state}
 \vec{\eta}_{q,j,\alpha}\coloneqq\vec{\theta}^{(q)}+\pi\alpha~\vec{\epsilon}_j\, ,
\end{align}
as initial conditions for the dynamics of Eq.~\eqref{eq:kuramoto_cycle}, with $\alpha\geq 0$. 
The parameter $\alpha$ is increased from zero to $\alpha_{q,j}$ which we define as the largest value such that $\vec{\eta}_{q,j,\alpha}$ converges back to $\vec{\theta}^{(q)}$ under the dynamics of 
Eq.~\eqref{eq:kuramoto_cycle}. 
The distance between the stable fixed point $\vec{\theta}^{(q)}$ and the boundary of its basin of attraction in the direction $\vec{\epsilon}_j$ is given by $\pi\alpha_{q,j}$. 

\begin{figure}
 \centering
 \includegraphics[width=.4\textwidth]{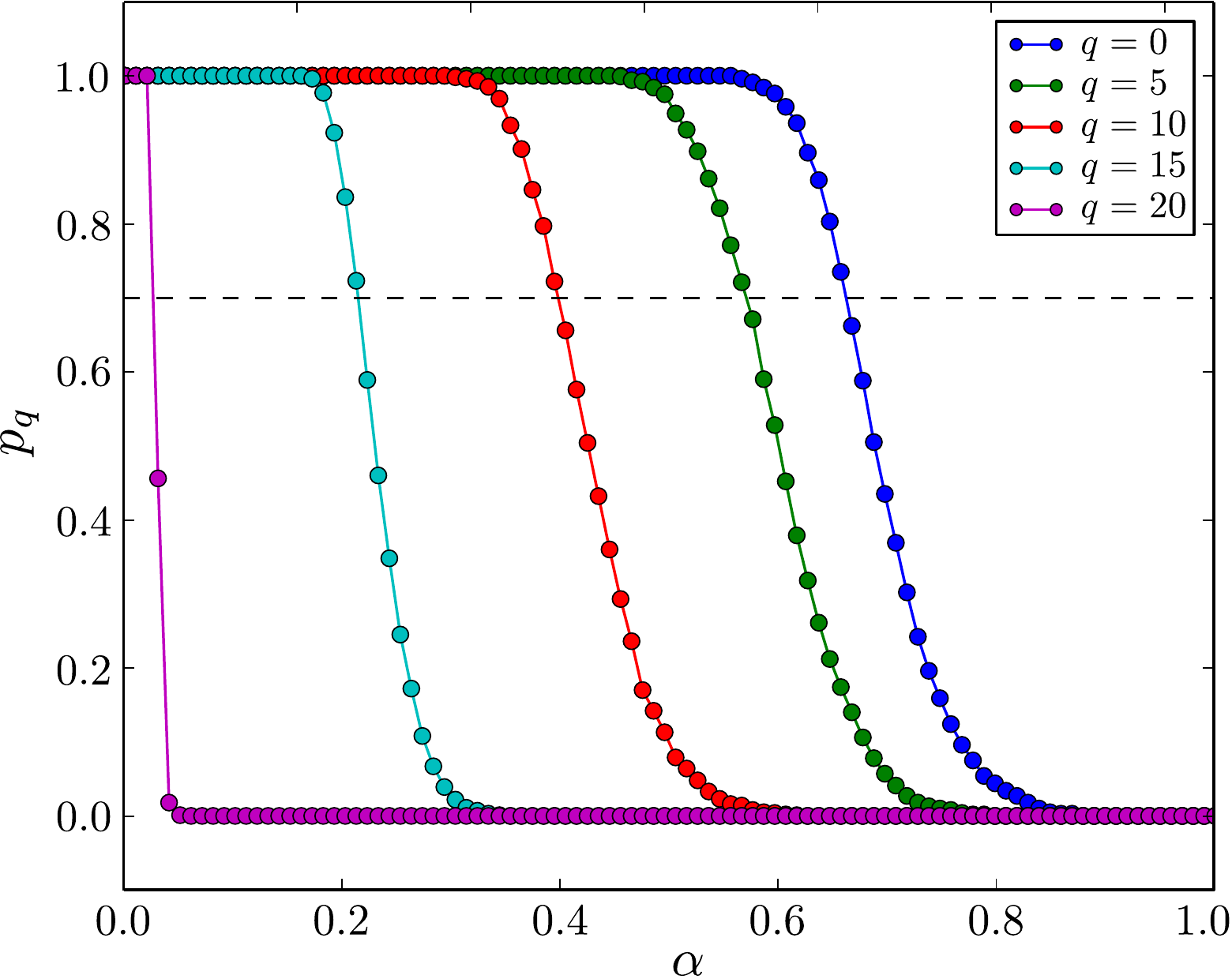}
 \caption{Proportion of perturbed states converging back to their reference stable fixed point with respect to the parameter $\alpha$ of Eq.~\eqref{eq:perturbed_state}, 
 for the equal-frequency Kuramoto model on a single cycle with $n=83$ nodes and winding numbers $q=0,5,10,15,20$ from right to left.}
 \label{fig:prop_alpha}
\end{figure}

We performed $4^{\text{th}}$-order Runge-Kutta simulations of the dynamics of Eq.~\eqref{eq:kuramoto_cycle} for $n=23,43,83,163,323$. 
These values are chosen to maximize the volume of the basin of attraction for the largest winding number $q_{\max}={\rm Int}(n/4)$, 
which, as mentioned above, vanishes when $n$ is a multiple of $4$.~\cite{Man16}
We took $d=1000$ randomly chosen perturbation directions and increased $\alpha$ by steps of $0.01$.
For each $q$, we can then estimate the proportion of the hypercube of side $\alpha$ centered at $\vec{\theta}^{(q)}$ which belongs to its basin of attraction as 
\begin{align}
 p_q(\alpha) &\coloneqq \frac{{\rm Card}\left\{\vec{\eta}_{q,j,\alpha} ~|~ \vec{\theta}(0) = \eta_{q,j,\alpha},\, \vec{\theta}(t\to\infty)=\vec{\theta}^{(q)}\right\}}{d}\, ,
\end{align}
where ${\rm Card}$ stands for the cardinality of the ensemble. 
In Fig.~\ref{fig:prop_alpha} we see that this proportion stays close to 1 for small values of $\alpha$ and quickly drops to zero around some $q$-dependent value of $\alpha$.
Given a threshold $\tau\in[0,1]$ we can then define 
\begin{align}\label{eq:alpha_tau}
 \alpha_{\tau}(q) &\coloneqq \sup\{\alpha~|~p_q(\alpha)\geq\tau\}\, ,
\end{align}
as a typical linear size of the basin of attraction. 
The abrupt drop of the curves in Fig.~\ref{fig:prop_alpha} implies that the precise value of $\tau$ is not too significant to understand the behavior of $\alpha_{q,j}$ with respect to $q$, 
provided that $\tau$ is neither too close to 1, nor to 0. 
We arbitrarily chose $\tau=0.7$, but checked that similar conclusions follow for $\tau=0.6$ and $0.8$. 

\begin{figure}
 \centering
 \includegraphics[width=.4\textwidth]{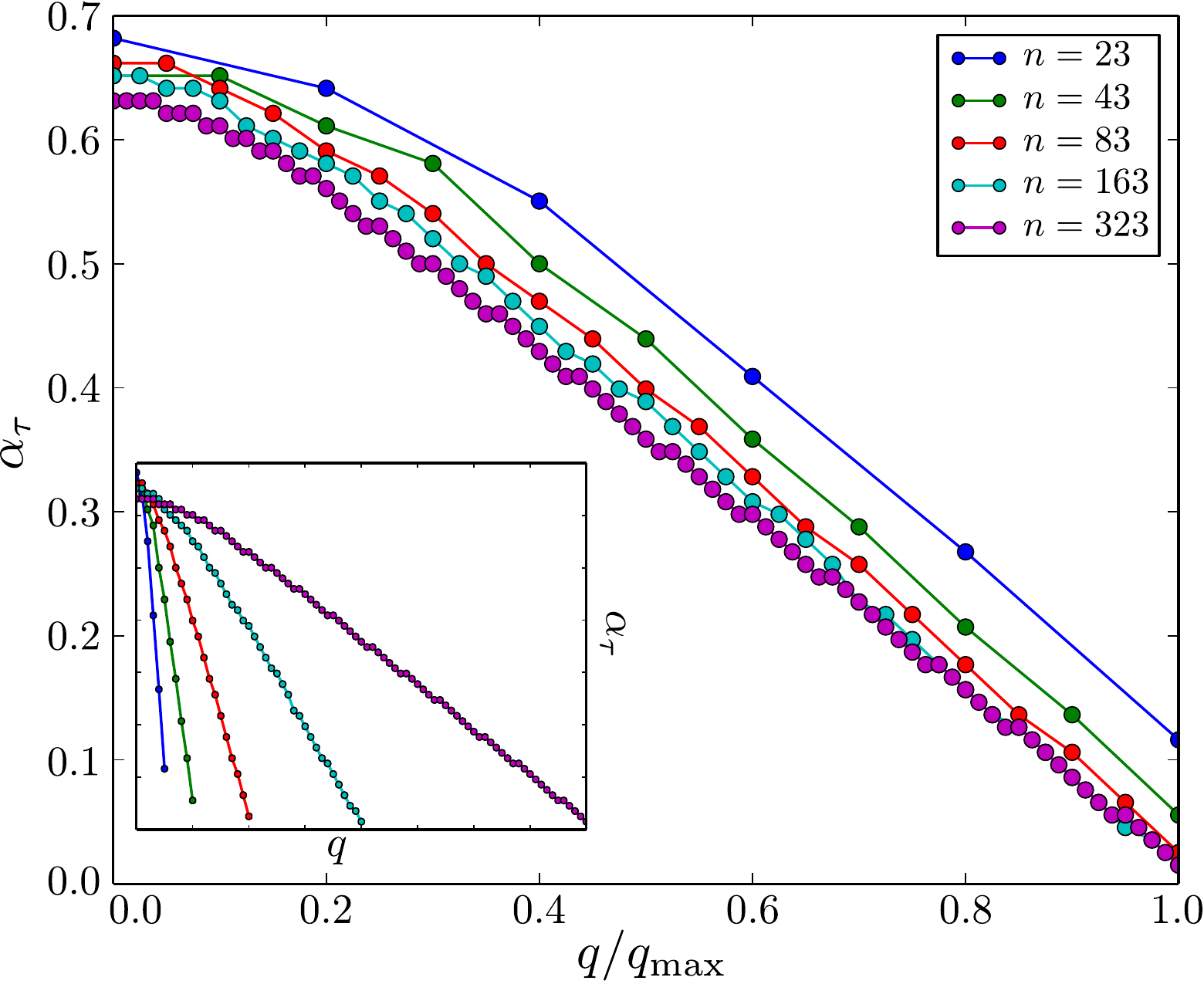}
 \caption{Typical linear size of the basins of attraction with respect to the winding number for the model of Eq.~\eqref{eq:kuramoto_cycle}. 
 Threshold values $\alpha_\tau$ defined such that $70\%$ of the $1000$ perturbed states $\vec{\eta}_{q,j,\alpha}$ converge to $\vec{\theta}^{(q)}$, 
 are plotted as a function of $q/q_{\max}$ (main panel) and $q$ (inset), for $n=23,43,83,163,323$.}
 \label{fig:dist_inf_num}
\end{figure}

\begin{figure}
 \centering
 \includegraphics[width=.4\textwidth]{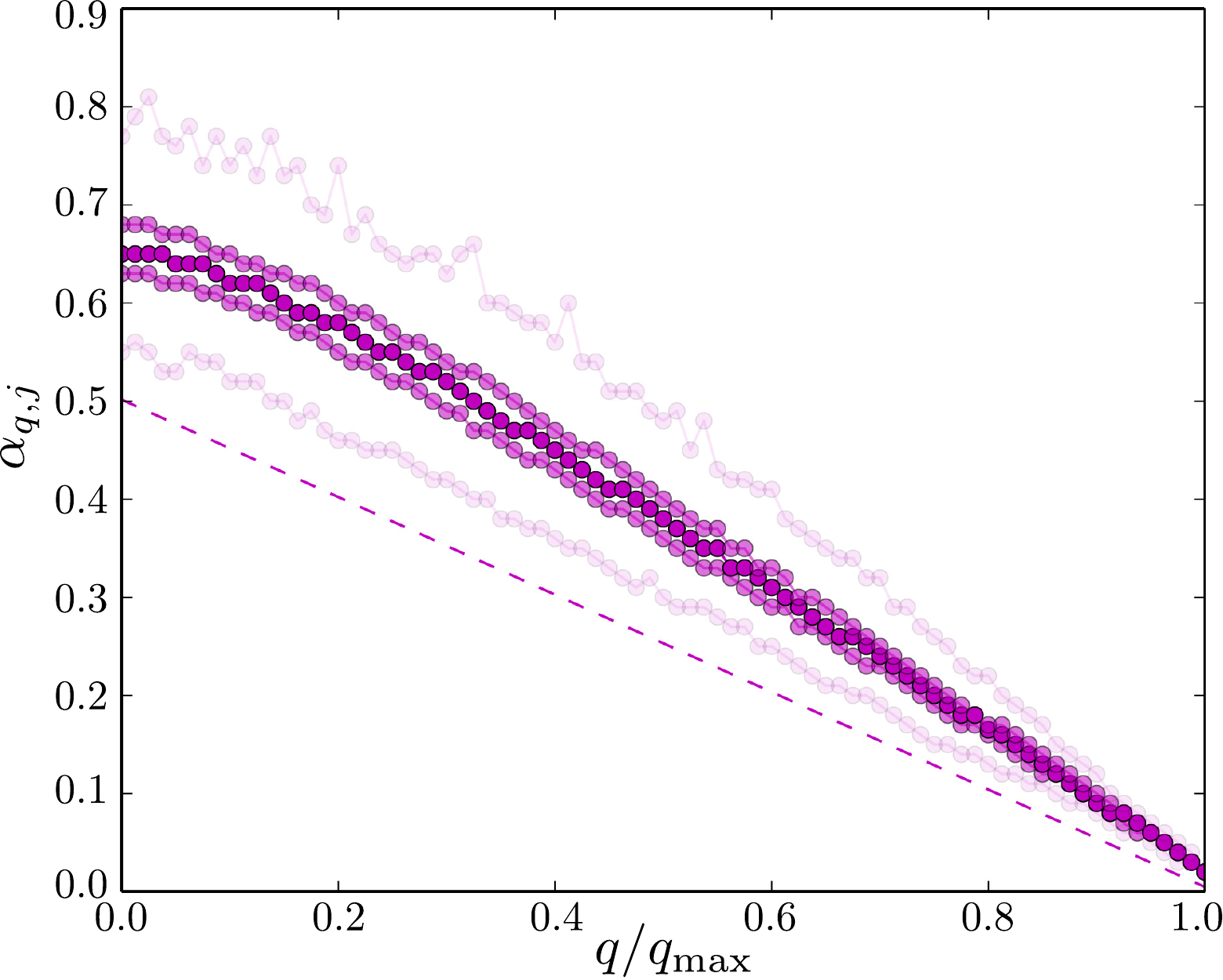}
 \caption{Quartiles of the values of $\alpha_{q,j}$ obtained from 1000 random directions $\vec{\epsilon}_j$ (purple dots), for $n=323$. 
 A quarter of the $\alpha_{q,j}$'s are between two vertically consecutive points. 
 Dashed line: distance between the stable fixed point $\vec{\theta}^{(q)}$ and the 1-saddles $\vec{\varphi}^{(q)}$ calculated in Eq.~\eqref{eq:distance}.}
 \label{fig:compar_theo_num}
\end{figure}

In Fig.~\ref{fig:dist_inf_num}, we plot $\alpha_\tau(q)$ for various system sizes. 
Except a saturation for small $q$'s, we observe a linear behavior of $\alpha_\tau$ with respect to $q$. 
Furthermore, curves for different values of $n$ varying by more than one order of magnitude are rescaled almost on top of one another when plotting them against $q/q_{\max}$. 
Both findings corroborate Eq.~\eqref{eq:distance}. 
Fig.~\ref{fig:compar_theo_num}, shows for each $q$, the quartiles of the values of $\alpha_{q,j}$ [defined by $\alpha_\tau(q)$ for $\tau=0.25,0.5,0.75$] and the extreme values 
$\min_j\alpha_{q,j}$ and $\max_j\alpha_{q,j}$, for $n=323$, 
as well as the distance between the stable fixed point $\vec{\theta}^{(q)}$ and the 1-saddle $\vec{\varphi}^{(q)}$ given by Eq.~\eqref{eq:distance} (dashed line).
All curves have linear behavior, except for small $q$. 
The discrepancy between numerics and Eq.~\eqref{eq:distance} comes from the fact that the random perturbations are not aligned with the direction of shortest distance to a 1-saddle. 
In other words, Eq.~\eqref{eq:distance} is a lower bound on the distance between the stable fixed point $\vec{\theta}^{(q)}$ and the boundary of its basin of attraction. 
As mentioned above, 1-saddles are closer to the stable fixed points than other saddle points, which are also on the boundary of the basins of attraction. 
Our method of using the 1-saddles to evaluate the volume of the basins of attraction underestimates it, but clearly gives its right parametric dependence in $n$ and $q$.  

\subsection{The size of the sync basin revisited}
The scaling obtained in Eq.~\eqref{eq:scaling}, for large values of $n$, is different from the Gaussian scaling postulated in Ref.~\onlinecite{Wil06}. 
The numerical method used there took initial conditions at random in the angle space $(-\pi,\pi]^n$, which would need a huge number of runs to reach a resolution allowing a fair 
estimate of the volumes of the basins of attraction. 
Even for a moderate resolution of $0.5$ in each angle direction, one would need approximately $(2\pi/0.5)^{83}\approx 10^{91}$ different initial condition, 
which is obviously unfeasible numerically. 
Hence, estimates based on brute-force numerical methods cannot catch the scaling behavior in dynamical systems with large dimensionality, 
especially for large winding numbers, which have very small basins of attractions. 

Our approach overcomes this difficulty.
Taking advantage of our knowledge of the stable fixed points, we are able to restrict the exploration of the basins of attraction to the neighborhood of the stable fixed points. 
We avoid scanning the whole angle space which significantly reduces the computation time and increase the accuracy of the method. 

\subsection{Non-identical frequencies}\label{sec:numeric_cyc_freq}
We introduced our method in the simplest case of a cycle network with identical frequencies. 
To generalize our understanding of the problem, we now add non-identical frequencies to the same cycle network. 
Even if we cannot obtain the stable fixed points analytically, we can find them numerically and then apply the same numerical procedure as in the identical frequency case. 
Instead of Eq.~\eqref{eq:kuramoto_cycle}, our single-cycle model is now defined by 
\begin{align}\label{eq:kuramoto_cycle_freq}
 \dot{\theta}_i &= P_i - K\sin(\theta_i-\theta_{i-1}) - K\sin(\theta_i-\theta_{i+1})\, ,
\end{align}
with $P_i$ randomly and homogeneously taken in $[-\beta,\beta]$, satisfying $\sum_iP_i=0$. 
For small values of $\beta$, the non-identical frequencies almost always lead to small variations of the fixed points,~\cite{Gil81}
and thus the volume of the basins of attraction should not change much. 
To find the stable fixed points of Eq.~\eqref{eq:kuramoto_cycle_freq}, we start with the fixed points for $\beta=0$ given in Eq.~\eqref{eq:angle_vec} and follow them with a $4^{\rm th}$-order 
Runge-Kutta implementation of Eq.~\eqref{eq:kuramoto_cycle_freq}, while gradually increasing $\beta$ to the desired value. 
This allows to identify and follow numerically the location of the stable fixed point $\vec{\theta}^{(q)}(\{P_i\})$, which is not anymore given by Eq.~\eqref{eq:angle_vec}, 
but is still characterized by its winding number $q$. 
We then perturb this stable fixed point in $1000$ random directions with increasing magnitude as in Eq.~\eqref{eq:perturbed_state} and 
apply the same procedure as in Sec.~\ref{sec:numeric_cyc_0} to evaluate the volume of the basins of attraction. 

\begin{figure}
 \centering
 \includegraphics[width=.4\textwidth]{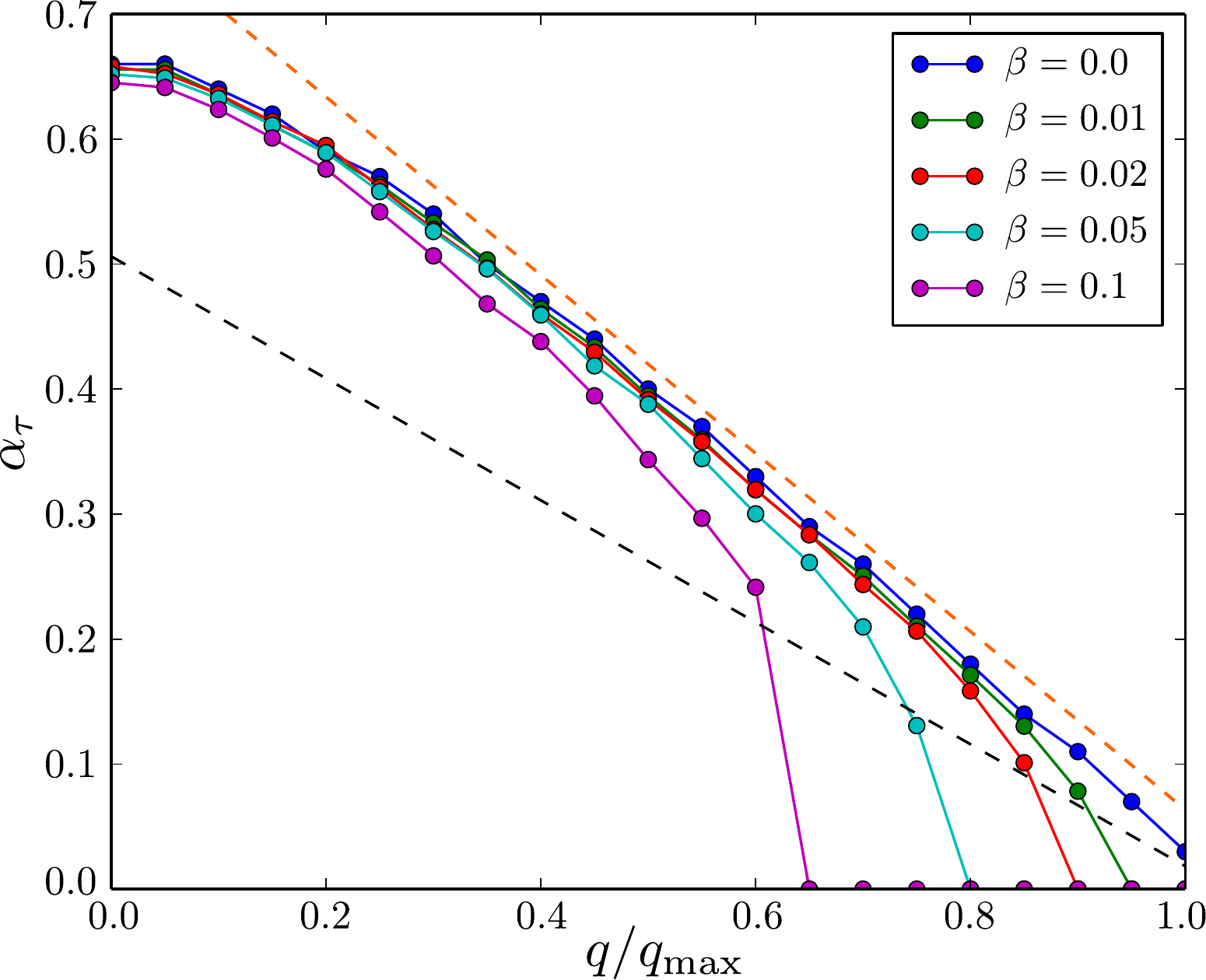}
 \caption{Typical linear size of the basins of attraction with respect to the winding number for the model of Eq.~\eqref{eq:kuramoto_cycle_freq}. 
 Threshold values $\alpha_\tau$ defined such that $70\%$ of the $1000$ perturbed states $\vec{\eta}_{q,j,\alpha}$ converge to $\vec{\theta}^{(q)}$, 
 are plotted as a function of $q/q_{\max}$ for $n=83$ and frequency distribution in $[-\beta,\beta]$ with $\beta=0,0.01,0.02,0.05,0.1$.
 The black dashed line is Eq.~\eqref{eq:distance} and the orange dashed line is a linear guide to the eye.}
 \label{fig:alpha_freqs}
\end{figure}

Results are shown in Fig.~\ref{fig:alpha_freqs} for $n=83$ and $\beta=0,0.01,0.02,0.05,0.1$.
For values of $q$ which are neither too small nor too large, the linear behavior of $\alpha_\tau$ is preserved, especially for small $\beta$.
As can be expected,~\cite{Col16b} as soon as we add some finite natural frequencies, the fixed points with large $q$ lose stability. 
More surprising, at first glance, is the abrupt drop of $\alpha_\tau$ for large $q$ with little change at small $q$. 
We offer an explanation for this behavior. 

\begin{figure*}
 \centering
 \includegraphics[width=.9\textwidth]{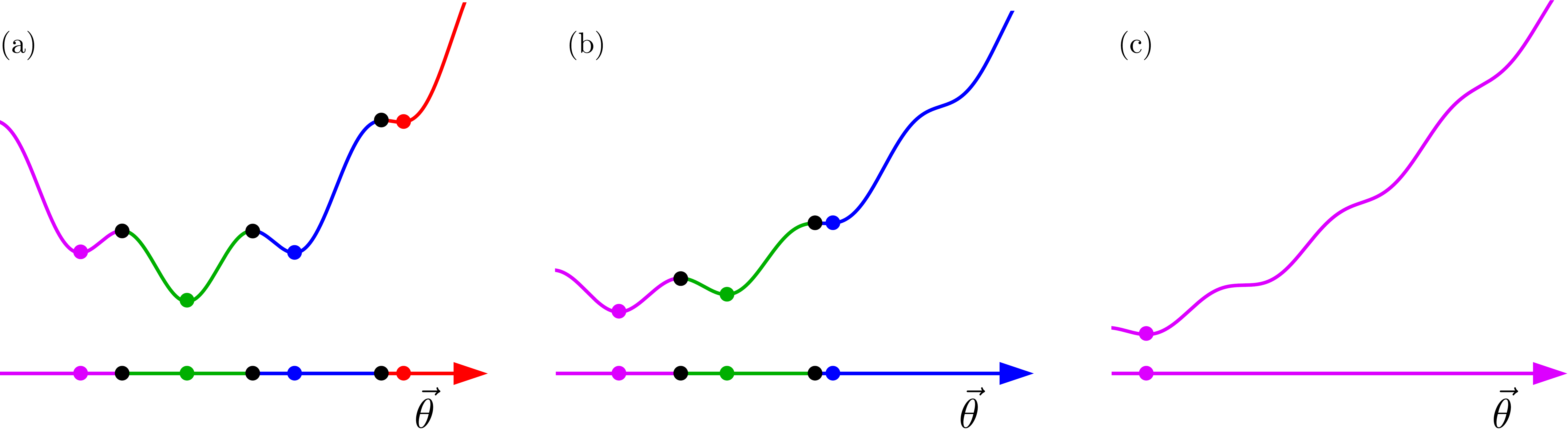}
 \caption{Schematic illustration of the projection of the Lyapunov function $\mathcal{V}$ of Eq.~\eqref{eq:lyapunov_function} in one dimension, with $P_i \in [-\beta,\beta]$. 
 The value of $\beta$ increases from left to right, leading to saddle-node bifurcations.
 Colored dots and lines are stable fixed points and their respective basins of attraction. Black dots are unstable fixed points. 
 Each stable fixed point gets closer to an unstable fixed point [panels (a) and (b)], but the volume of the basins of attraction does not change much unless one fixed point loses 
 stability [panel (c)].
 Fixed points with large winding numbers [red region in panel (a)] lose stability before fixed points with lower winding numbers (blue and green regions).}
 \label{fig:saddle-node}
\end{figure*}

The dynamics of Eq.~\eqref{eq:kuramoto_gen} is given by the gradient of the Lyapunov function
\begin{eqnarray}
\mathcal{V}(\vec{\theta})&=&-\sum_iP_i\theta_i-\sum_{i<j}K_{ij}\cos(\theta_i-\theta_j)~, \label{eq:lyapunov_function}\\
-\frac{\partial \mathcal{V}}{\partial \theta_i} &=& \dot{\theta}_i\, .
\end{eqnarray}
Increasing $\beta$ modifies $\mathcal{V}$ [the first term on the right-hand side of Eq.~\eqref{eq:lyapunov_function}] and makes the fixed points move in angle space. 
Eventually, a stable fixed point $\vec{\theta}^{(q)}$ will meet an unstable fixed point and then lose stability through a saddle-node bifurcation. 
In Fig.~\ref{fig:saddle-node}, we give a schematic illustration of $\mathcal{V}$ on a cycle, projected on an appropriate direction in angle space, such that stable and unstable 
fixed points are aligned in one angle dimension.
As long as a fixed point remains stable, the volume of its basin of attraction does not change much [compare the green segments in Figs.~\ref{fig:saddle-node}(a) and \ref{fig:saddle-node}(b)].
The fixed point then abruptly vanishes when $\beta$ becomes too large.
Since $\alpha_\tau$ is an average over many randomly chosen directions, its value abruptly drops when the stable fixed point vanishes. 

In Appendix~\ref{ap:max_winding}, we furthermore estimate the maximal winding number possible for a given width $\beta$ of frequencies distribution 
and find that it agrees qualitatively with the numerically observed maximal winding numbers. 

\section{Meshed Networks}\label{sec:algo}
We finally extend the perturbation method described above to more complicated, meshed networks. It is a two-stage method where we 
first numerically identify fixed points of Eq.~(\ref{eq:kuramoto_gen}) on complex graphs and
second perturb the obtained stable fixed points in the same way as in Sec.~\ref{sec:single_cycle}. 

\subsection{Identifying stable fixed points}
Stable fixed points are much harder to find on complex graphs. 
Except for the $\vec{\theta}^{(0)}=(0,...,0)$ fixed point for equal frequencies, they are usually impossible to find analytically. 
To tackle this problem, we construct a numerical algorithm similar to but different from the one proposed in Ref.~\onlinecite{Man16}.

\begin{figure*}
\centering
\subfigure{
    \includegraphics[width=.315\textwidth]{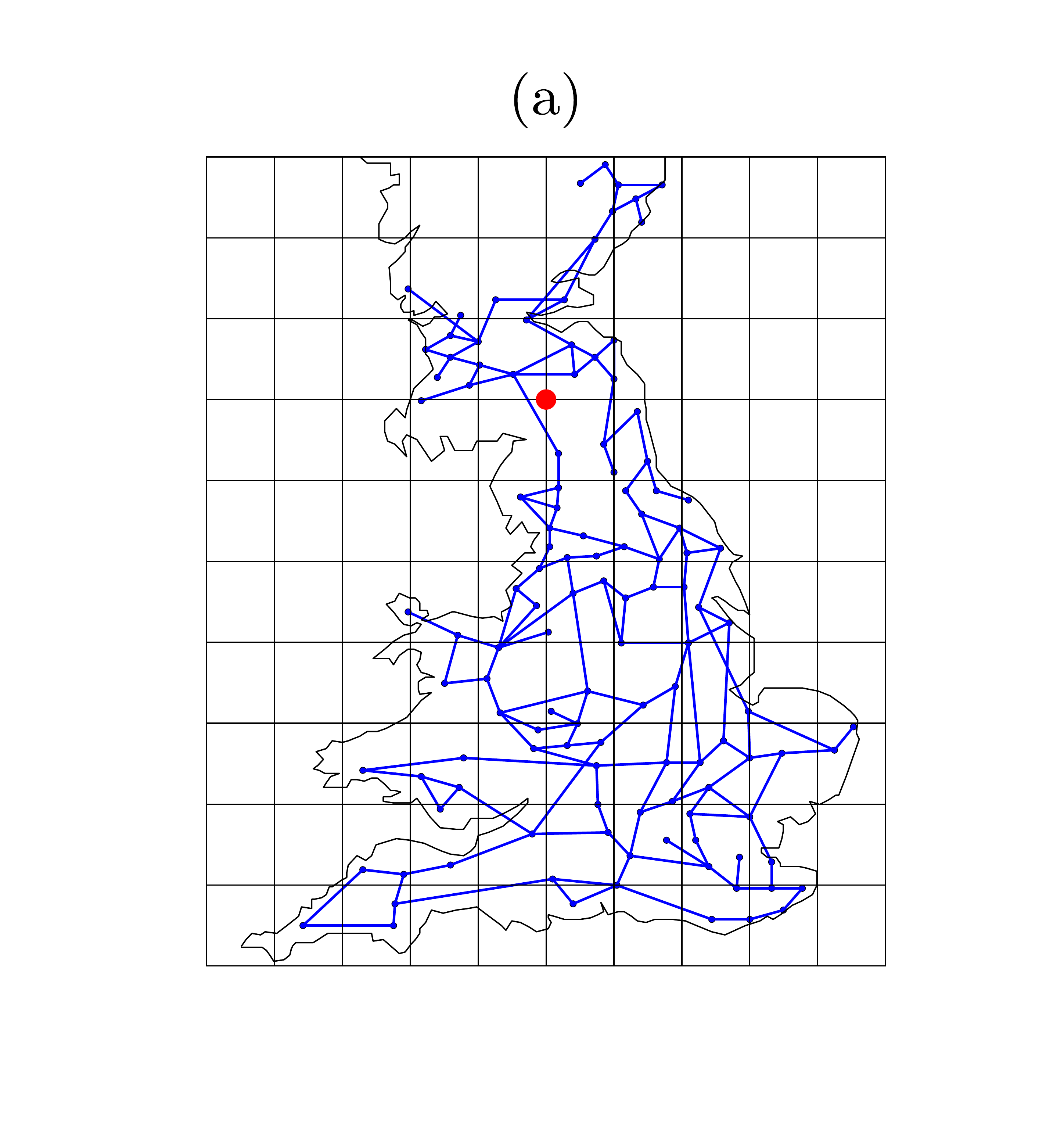}
    \label{sfig:UK_geo_em}
}
\subfigure{
 \includegraphics[width=.315\textwidth]{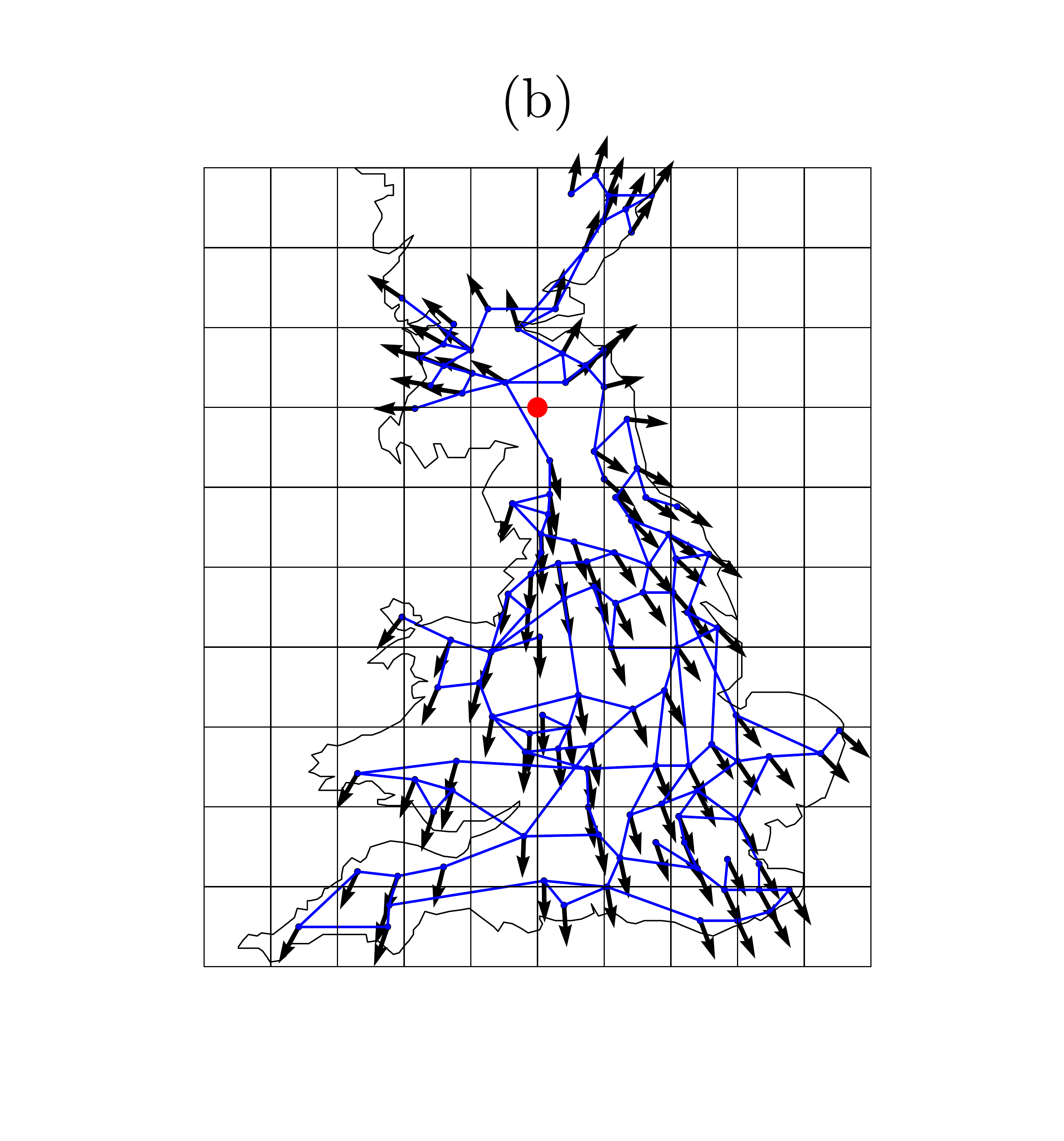}
 \label{sfig:fina_UK}
}
\subfigure{
  \includegraphics[width=.315\textwidth]{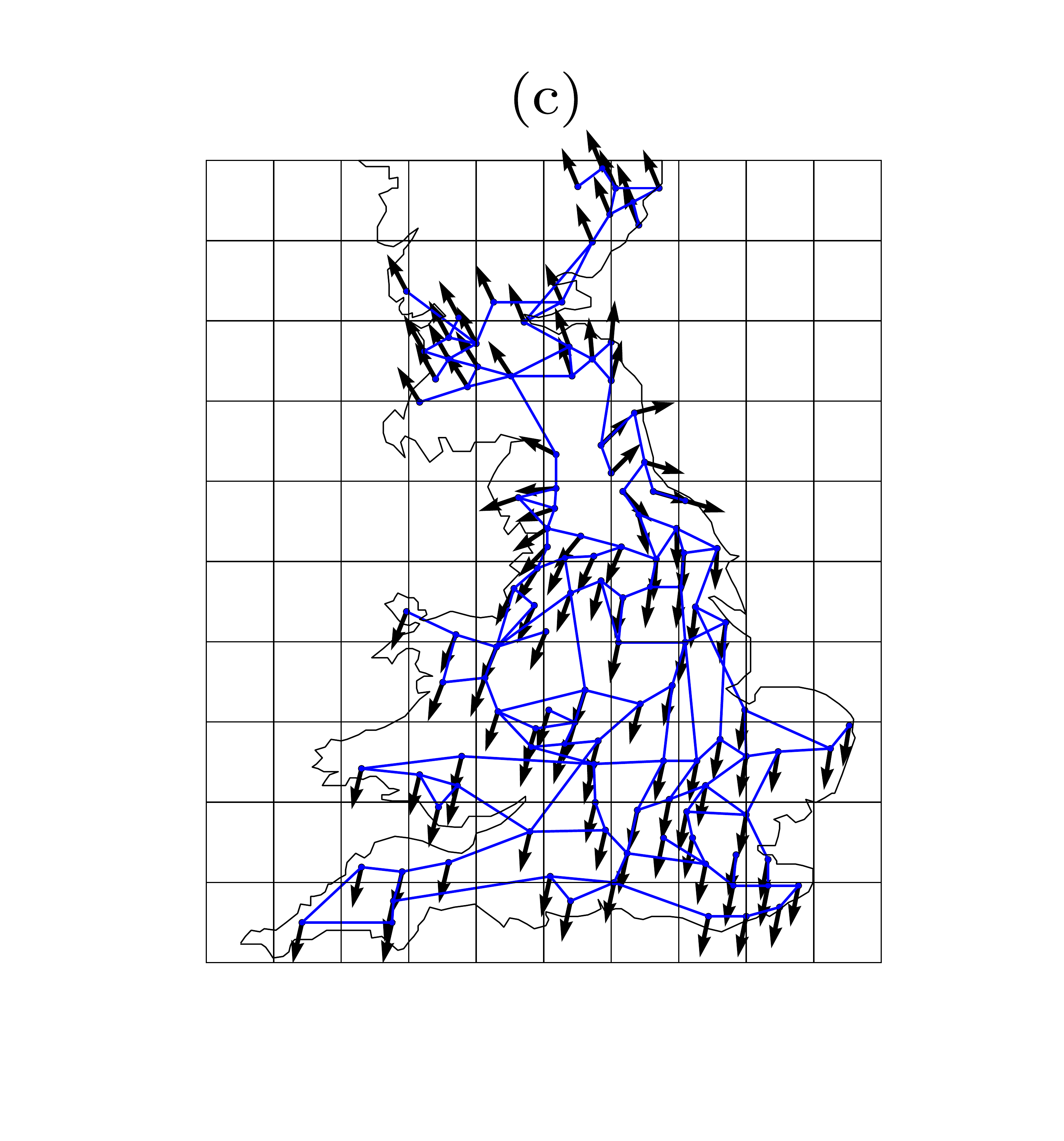}
  \label{sfig:init_UK}
}
\vspace*{-10mm}
\caption{(a): Geographic embedding of the UK high voltage grid with a square lattice. 
 (b): Initial condition with a $q=1$ vortex centered at the red dot. 
 (c): Stable fixed point towards which the initial state of panel (b) converges under the dynamics of Eq.~(\ref{eq:kuramoto_gen}) with $P_i=0$, $\forall i$.}
\label{fig:UK_graph_init_fina}
\end{figure*}

From Refs.~\onlinecite{Del16,Dor13}, we know that two fixed points of Eq.~\eqref{eq:kuramoto_gen} differ only by a collection of loop flows quantized by their winding numbers 
(similar to vortices in superconductors). 
We define a vector composed of the winding numbers on each cycle of a graph ${{G}}$ as 
\begin{eqnarray}
\vec{q}_{{{G}}}(\vec{\theta})=(q_1,q_2,...,q_m)\, ,
\end{eqnarray}
where $m$ is the total number of cycles in ${{G}}$. 
Assuming $|\theta_i-\theta_j|<\pi/2$ for all connected nodes $i$, $j$, each stable fixed point $\vec{\theta}^*$ can be uniquely labelled by its winding vector 
$\vec{q}_{{{G}}}(\vec{\theta}^*)$.~\cite{Del16,Dor13,Man16}
For $P_i\equiv0$, the Lyapunov function, Eq.~\eqref{eq:lyapunov_function}, reduces to 
\begin{align}
\mathcal{V}(\vec{\theta})&=-\sum_{i<j}K_{ij}\cos(\theta_i-\theta_j)~, \label{eqxy}
\end{align}
which is the Hamiltonian of a XY model, describing the interaction of planar classical spins.\cite{Faz01} 
Stable fixed points are the local minima of this energy function and it is known that they correspond to vortex-carrying states, 
i.e. states with non-zero winding vector $\vec{q}_{G}(\vec{\theta})$. 
We therefore search for stable fixed points via an iterative process starting from vortex-carrying initial states described by
\begin{equation}\label{eqtan}
\theta_i=q\arctan\left(\frac{y_i-y_0}{x_i-x_0}\right)\, ,
\end{equation}
where $(x_0,y_0)$ are the coordinates of the center of the vortex with charge/vorticity $q \in \mathbb{Z}$, and $(x_i,y_i)$ the coordinates of the position of $\theta_i$.
To find stable fixed points, the algorithm reads: 

\begin{enumerate}
\item Define a two-dimensional embedding of the network. Use this to superimpose a regular lattice of coordinates
on the network. This is shown in Fig.~\ref{sfig:UK_geo_em}. 
\item Set $(x_0,y_0)$ to a node of the regular lattice.
\begin{enumerate}
\item Using Eq.~(\ref{eqtan}), define a new initial state.
\item Follow numerically Eq.~(\ref{eq:kuramoto_gen}) on the considered meshed graph until a stable fixed point is reached.
\item Each stable fixed point can be unambiguously identified by its winding vector. Use this to determine if the fixed point just found is a new one. If yes, store it.
\end{enumerate}
\item Go back to step 2. 
\end{enumerate}

As complex meshed network, we consider the UK high voltage grid which is composed of 120 nodes and 165 lines. To illustrate the algorithm just described, 
Figs.~\ref{sfig:fina_UK} and \ref{sfig:init_UK} show an initial condition and the stable state toward which it dynamically converges respectively. 
Stable fixed points with many vortices are obtained by setting $|q|$ to large values, in our case $q\in \{-50, ... ,50\}$. 
The dynamics will then split this initial vortex into several vortices with smaller $q$'s, located on different cycles of the network. 
This method can be used on any network whether complex or regular. 
Time evolving Eq.~(\ref{eq:kuramoto_gen}) on the UK grid with this initial condition returns only stable fixed points. In this way
we found more than 4000 different stable fixed points of Eq.~(\ref{eq:kuramoto_gen}) with $P_i=0$.

\subsection{Estimating the volume of basins of attraction}\label{sec:meshed}
Having identified stable fixed points $\vec{\theta}^{(\vec{q})}$ of Eq.~\eqref{eq:kuramoto_gen} on the UK grid, 
we next follow the same procedure as in Sec.~\ref{sec:numeric_cyc_0}, 
and measure the volume of their basins of attraction. 
\begin{figure}
 \centering
 \includegraphics[width=.36\textwidth]{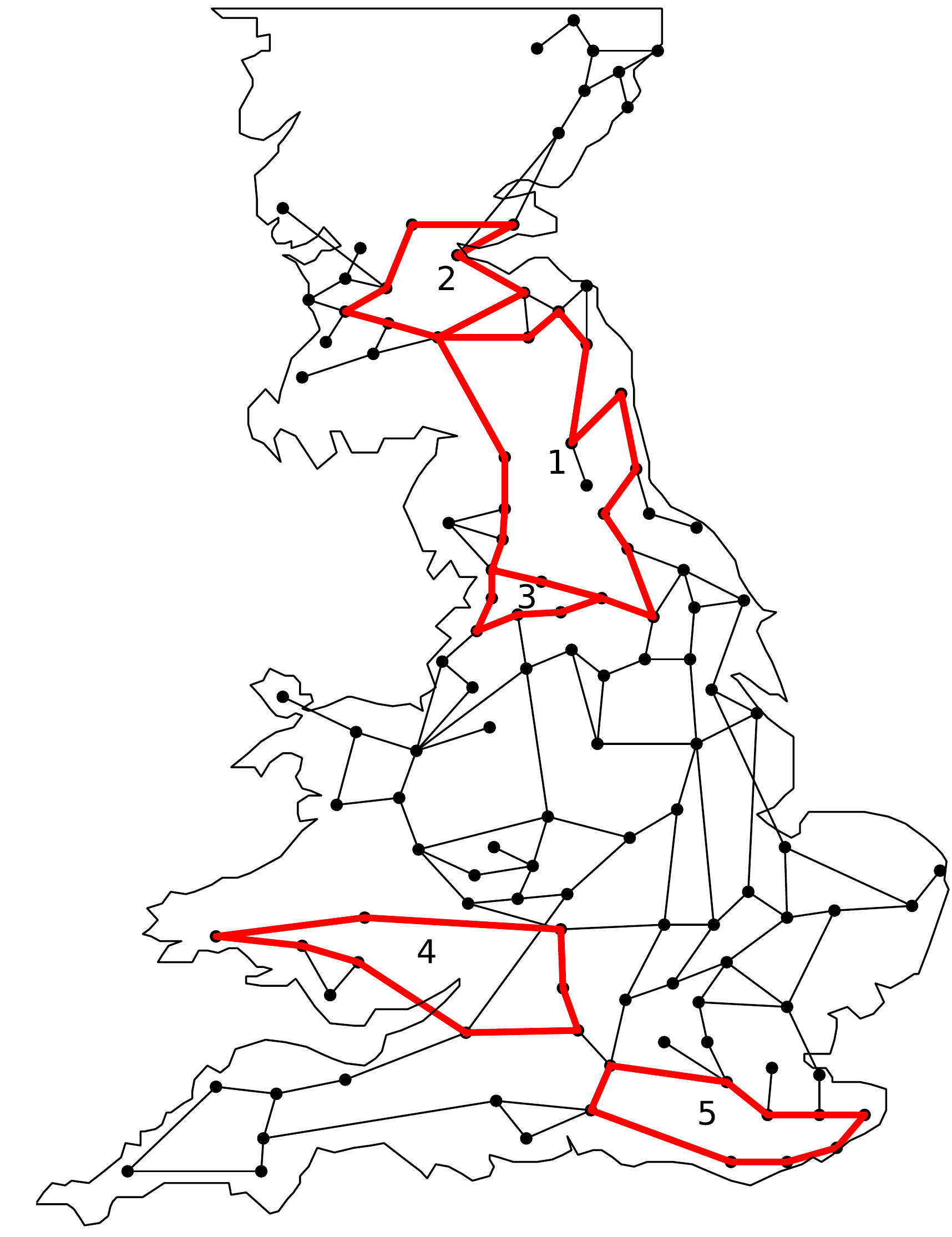}
 \caption{High voltage UK AC transmission grid used as meshed network. The five cycles we focus on are indicated in red. Note that cycle 4 is traversed but not interrupted by an edge.}
 \label{fig:UK_graph}
\end{figure}

We focus on stable fixed points with non-zero winding number only on the five cycles in red in Fig.~\ref{fig:UK_graph}. 
We introduce a shorthand notation with the winding numbers of these cycles only
\begin{eqnarray}
\vec{q}_{\rm sh}=(q_{1},q_{2},q_{3},q_{4},q_{5}).
\end{eqnarray}
Taking each cycle independently, the maximum winding numbers are ${q}_{1}^{\rm max}=4$, ${q}_{2}^{\rm max}=2$, ${q}_{3}^{\rm max}=1$, ${q}_{4}^{\rm max}=2$, 
${q}_{5}^{\rm max}=2$. 

\begin{figure*}
 \centering
 \includegraphics[width=0.8\textwidth]{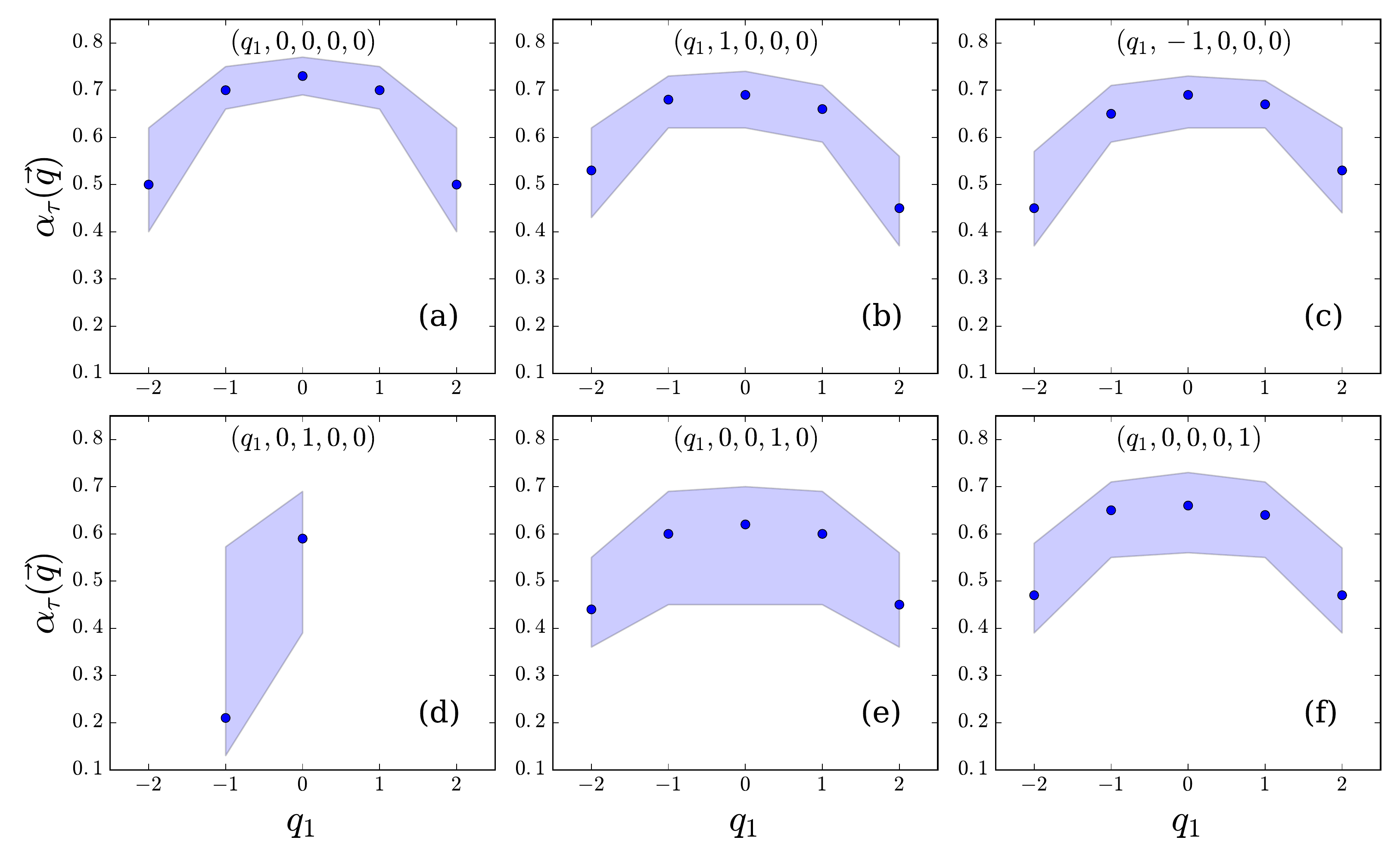}
 \caption{Median (blue dots) and interquartiles (blue areas) of the values of $\alpha_{\vec{q},j}$ obtained from 1000 random perturbations $\vec{\epsilon}_j$ 
 of fixed points of Eq.~(\ref{eq:kuramoto_gen}) with $P_i=0$, $\forall i$ on the UK grid of Fig.~\ref{fig:UK_graph} as a function of the winding number on cycle 1 ($q_1$).}
 \label{fig:alpha_c_UK}
\end{figure*}
In Fig.~\ref{fig:alpha_c_UK}, we show $\alpha_\tau$ of Eq.~(\ref{eq:alpha_tau}) for $\tau=0.5$ and the interquartile values for various stable fixed points identified by their 
unique combination of winding numbers. 
For $P_i=0$, Eq.~(\ref{eq:kuramoto_gen}) is symmetric under $\theta_i\rightarrow -\theta_i$, $\forall i$. 
This implies that the volume of each basin of attraction is invariant under $\vec{q}_{G}\rightarrow -\vec{q}_{G}$. 
This symmetry can be seen on Fig.~\ref{fig:alpha_c_UK}(a) which is symmetric under $q_1\rightarrow -q_1$, as there is only one cycle with a non-zero winding number. 
Alternatively, $\vec{q}_G\to-\vec{q}_G$ interchanges panels (b) and (c) with $q_1\to-q_1$. 
For fixed points with more than one cycle carrying a non-zero winding number, there is no symmetry for $q_1 \rightarrow -q_1$ 
[see Figs.~\ref{fig:alpha_c_UK}(b)--(f)] and the asymmetry is even more significant if the vortices are close to each other [see Figs.~\ref{fig:alpha_c_UK}(b) and \ref{fig:alpha_c_UK}(d)], 
because there are few intermediate nodes which can screen the effect of one vortex on the other and so the cycles interact strongly.

\begin{figure}
 \centering
 \includegraphics[width=.2\textwidth]{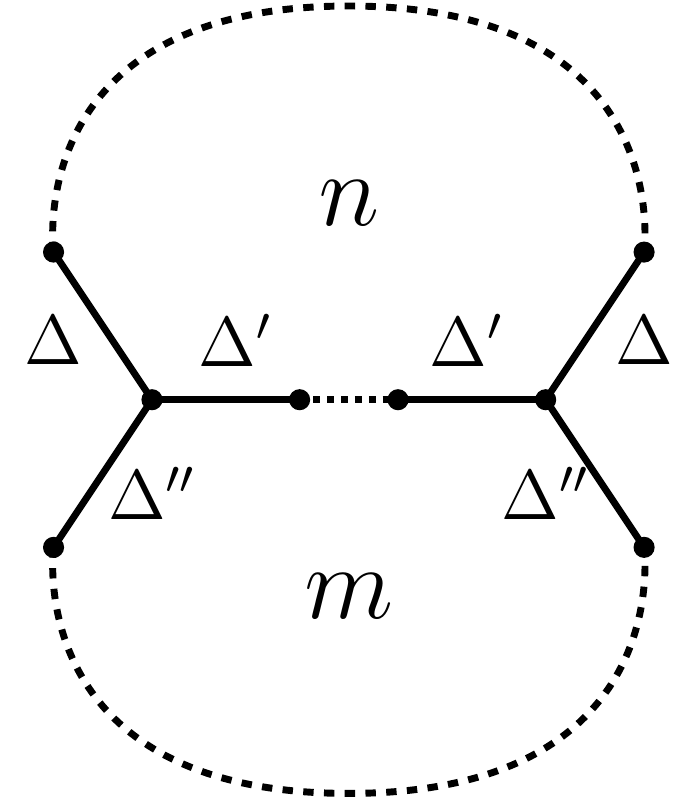}
 \caption{Two cycles sharing $\ell$ edges. The top cycle has $n$ nodes and the bottom one $m$ nodes. 
 Angle differences are given by $\Delta$, $\Delta'$ and $\Delta''$ for Eq.~(\ref{eq:kuramoto_gen}) with equal frequencies.}
 \label{fig:cycles_UK}
\end{figure}

In Fig.~\ref{fig:alpha_c_UK}(d), we did not find any fixed point with winding vector $\vec{q}_{\rm sh}=(1,0,1,0,0)$ but we found one with $\vec{q}_{\rm sh}=(-1,0,1,0,0)$. 
To understand this, we consider the simplified situation of two connected cycles of size $m$ and $n$, sharing $\ell$ edges as depicted on Fig.~\ref{fig:cycles_UK}. 
We consider the case when $q^{(n)}=1$ and $q^{(m)}=0$. 
At a fixed point, Eq.~(\ref{eq:kuramoto_gen}) implies
\begin{equation}\label{eq:flux}
\sin(\Delta)=\sin(\Delta')+\sin(\Delta'')~,
\end{equation}
and Eq.~(\ref{eq:winding}) gives 
\begin{equation}\label{eq:wind_cy}
(m-\ell)\cdot\Delta''-\ell\cdot\Delta'=2\pi q^{(m)}=0\, .
\end{equation}
From Ref.~\onlinecite{Del17}, if $\ell>1$, we have $|\Delta|$,$|\Delta'|$,$|\Delta''|\leq\pi/2$. 
Eq.~(\ref{eq:flux}) and Eq.~(\ref{eq:wind_cy}) imply that $\Delta >\Delta'$.
Therefore, to have $q^{(n)}=1$, we must have $\Delta > {2\pi}/{n}$. 
Thus, if we add edge-sharing cycles with zero winding number to a main cycle with a non-zero winding number, some of the angle differences must increase. 
When $\Delta$ is large, this can bring $\Delta > {\pi}/{2}$ where stability is lost.\cite{Del17} 
Adding a cycle carrying a non-vanishing winding number makes the situation even more critical. 
If we isolate cycles 1 and 3, which correspond to Fig.~\ref{fig:cycles_UK} with $n=16$, $m=7$ and $\ell=2$, 
an easy calculation shows that there exist stable fixed points with $({q^{(16)}},q^{(7)})=(1,1)$ and $({q^{(16)}},q^{(7)})=(1,-1)$. 
However, when we consider the complete network, only the solution with $({q^{(16)}},q^{(7)})=(1,-1)$ remains stable. 
This comes from the fact that, when both winding numbers have the same signs, the angle differences on the shared edges benefit only to one of the cycles. 
The other cycle then has to make a winding number out of a reduced number of edges, implying larger angle differences. 
Finally, when we take the complete network, we put cycles next to the two initial ones and make the angle differences even larger, until  stability is lost. 
When the winding numbers have opposite signs, both cycles benefit from the angle differences on the shared edges, 
which leads to smaller angle differences than in the previous case.  This explains why the fixed point 
$\vec{q}_{\rm sh}=(-1,0,1,0,0)$ is stable while $\vec{q}_{\rm sh}=(1,0,1,0,0)$ is not.

To conclude this section, we note that a meshed network has an effect similar to 
the case considered in Sec.~\ref{sec:numeric_cyc_freq} with $P_i\neq 0$ in that, compared to the single-cycle network, (i) there are fewer stable
fixed points with large winding numbers and (ii) the volume of basins of attraction of fixed points with small winding numbers seem to be unaffected.

\section{Conclusion}\label{sec:conclusion}
We have developed a numerical method to investigate the volume of basins of attraction of fixed points in dynamical systems. 
Our method first locates the stable fixed points of the dynamical system, using an algorithm based on the concept of loop flows.~\cite{Del16} 
Second, it pertubs them in random directions with increasing magnitude. 
The proportion of pertubed states that converge back to the initial fixed point allows to evaluate the radius of the basin of attraction and then its volume. 

We then used our method to investigate the Kuramoto model on a cycle with identical frequencies. 
We obtained that the volume of the basin of attraction is proportional to $(1-4q/n)^n$, contrasting with the Gaussian distribution suggested in Ref.~\onlinecite{Wil06}. 
We then extended the application of our method to the Kuramoto model on a cycle with non-identical frequencies and to the Kuramoto model on meshed networks. 
These two generalizations render the investigations of the basins of attraction much less tractable, which imposes to rely on numerics. 
We believe that our method significantly speeds up these investigations. 

Compared to other existing methods to investigate basins of attraction, our method has three main advantages:
\begin{itemize}
 \item It does not require a Lyapunov function of the dynamical system considered,~\cite{Gen85,Naj16} which is complicated to find in general;
 \item It is not limited to quadratic or polynomial systems;~\cite{Ama07}
 \item The investigation is guided by our knowledge of the system and avoids to randomly pick initial conditions in the state space.~\cite{Wil06,Men13}
\end{itemize}
These advantages come with the drawback that we limit our investigations to the volume of the basins of attraction and have no indications about their shape. 
In particular, our method is probably not adapted to the investigation of fractal basins of attraction.~\cite{Gre87}
We think that our method may be useful in many other contexts including finding local energy minima in planar spin glasses~\cite{Mez87}  
and disordered Josephson junction arrays~\cite{Faz01} among others.

\section*{Acknowledgment}
This work has been supported by the Swiss National Science Foundation under an AP Energy Grant.
We thank Jimmy Dubuisson for discussions at the early stage of this project.

\begin{appendices}
\numberwithin{equation}{section}

\section{Angle vector of the 1-saddle points}\label{ap:1saddle}
We give here details of the computation of the components of the angle vector $\vec{\varphi}^{(q')}\in{\cal H}_{n-1}$ defined in Sec.~\ref{sec:analytic}.
The vector $\vec{\varphi}^{(q')}$ is the 1-saddles on a cycle of length $n$, 
with winding number $q'$ and the only angle difference exceeding $\pi/2$ located on the edge between vertices $k-1$ and $k$.
Its components are given by
\begin{align}
 \varphi_i^{(q')} &= \left\{
 \begin{array}{ll}
  i\Delta' - S_k\, , & \text{if }i<k\, ,\\
  (i-2)\Delta' + \pi - S_k\, , & \text{if }i\geq k\, ,
 \end{array}
 \right.
\end{align}
where $S_k$ is a constant angle shift guaranteeing that the sum of components is zero,
\begin{align}
S_k \coloneqq n^{-1}\left[\sum_{j=0}^{k-1}j\Delta' + \sum_{j=k}^{n-1}\left((j-2)\Delta' + \pi\right)\right]\, .
\end{align}
Some algebra gives 
\begin{align}
 \varphi_i^{(q')} &= \pi\left[\frac{2q'-1}{n-2}i+\frac{-2n^2k+2nk-8q'k-n}{2n(n-2)} + T_{i}^{(k)}\right]\, ,
\end{align}
where
\begin{align}
 T_{i}^{(k)} &= \left\{
  \begin{array}{ll}
   \frac{10nq'-n^2}{2n(n-2)}\, , & \text{if }i<k\, ,\\
   \frac{2nq'+n^2}{2n(n-2)}\, , & \text{if }i\geq k\, .
  \end{array}
  \right.
\end{align}

\section{Maximal winding number on a cycle with random natural frequencies}\label{ap:max_winding}
According to Ref.~\onlinecite{Del16}, the angle difference on the edge between vertices $k$ and $k+1$ is given by
\begin{align}\label{eq:arcsine}
 \Delta_{k,k+1} &= \arcsin(\varepsilon_q+ P_{k,k+1}^*/K)\, ,
\end{align}
where $ P_{k,k+1}^*\coloneqq\sum_{j=1}^k P_j$ is a reference flow and $\varepsilon_q$ is the loop flow parameter determining the winding number of the fixed point. 
For the sake of simplicity we take $K=1$. 
The frequencies $P_k$ are taken randomly and homogeneously in the interval $[-\beta,\beta]$. 
Their expectation and variance are 
\begin{align}
 \mathbb{E}(P_k) &= 0 &\text{and}& &{\rm var}(P_k) &= \beta^2/3\, .
\end{align}
Expectation and variance for the $ P_{k,k+1}^*$ are then
\begin{align}
 \mathbb{E}(P_{k,k+1}^*) &= 0 &\text{and}& &{\rm var}( P_{k,k+1}^*) &= k\cdot\beta^2/3\, .
\end{align}
For $n$ sufficiently large, we then expect typical excursions of magnitude $\sqrt{k}\beta/\sqrt{3}$ of $ P_{k,k+1}^*$ away from its average $\mathbb{E}=0$. 
It is known~\cite{Del16} that on cycles with finite natural frequencies, stable fixed points may have one angle diffence slightly larger than $\pi/2$ before losing stability at $\pi/2+\delta$. 
As $\delta$ is always small, we will approximate the loss of stability to happen when $\Delta_{k,k+1}=\pi/2$, i.e. when the argument of the arcsine in Eq.~\eqref{eq:arcsine} is equal to one.
Finally, we approximate $\varepsilon_q$ by its value when $\beta=0$,
\begin{align}
 \varepsilon_q &\approx \sin(2\pi q/n)\, .
\end{align}

\begin{table}
 \begin{tabular}{|c|c|c|}
  \hline
  $\beta$ & $q_{\max}(\beta)$: Eq.~\eqref{eq:est_qmax} & $q_{\max}(\beta)$: Sec.~\ref{sec:numeric_cyc_freq}  \\
  \hline
  0 & 20.75 & $\{20,20,20,20,20\}$ \\
  0.01 & 17.14 & $\{17,17,18,18,18\}$ \\
  0.02 & 15.62 & $\{16,16,17,17,17\}$ \\
  0.05 & 12.56 & $\{13,14,14,15,15\}$ \\
  0.1 & 8.97 & $\{10,11,12,12,13\}$ \\
  \hline
 \end{tabular}
 \caption{Maximal values of $q$ with respect to $\beta$, estimated by Eq.~\eqref{eq:est_qmax} and 
 obtained with 5 random frequency distributions following the simulation process of Sec.~\ref{sec:numeric_cyc_freq}.}
 \label{tab:q_max}
\end{table}

Putting everything together and taking $k$ to be the average value of the indices, i.e. $k=n/2$, Eq.~\eqref{eq:arcsine} gives 
\begin{align}
 \sin(2\pi q_{\max}/n) + \sqrt{n}\beta/\sqrt{6} &= 1\, ,
\end{align}
which gives a maximal possible value of $q$ before losing stability, with respect to $\beta$,
\begin{align}\label{eq:est_qmax}
 q_{\max}(\beta) &= (2\pi)^{-1}n\arcsin(1-\sqrt{n}\beta/\sqrt{6})\, .
\end{align}
Simulated and estimated values of $q_{\max}$ are given in Table~\ref{tab:q_max} for various values of $\beta$. 
Even if the simplifications assumed to obtain Eq.~\eqref{eq:est_qmax} underestimates $q_{\max}$, it is in fair agreement with numerical obtained values.

\section{Distance between stable fixed points and saddle points}\label{ap:fixed_point_dist}
We justify numerically the two statements of Sec.~\ref{sec:analytic}, that the 1-saddles are the closest unstable fixed points to the stable fixed points, 
and that stable fixed points and 1-saddles are the closest if they have the same winding number. 
We write $\vec{\varphi}^{(q',\ell)}_i$ for the unstable fixed point with winding number $q'$ and $\ell$ angle differences larger than $\pi/2$, 
where the index $i$ labels the different fixed points with same $q'$ and $\ell$.

\begin{figure}
 \centering
 \includegraphics[width=.4\textwidth]{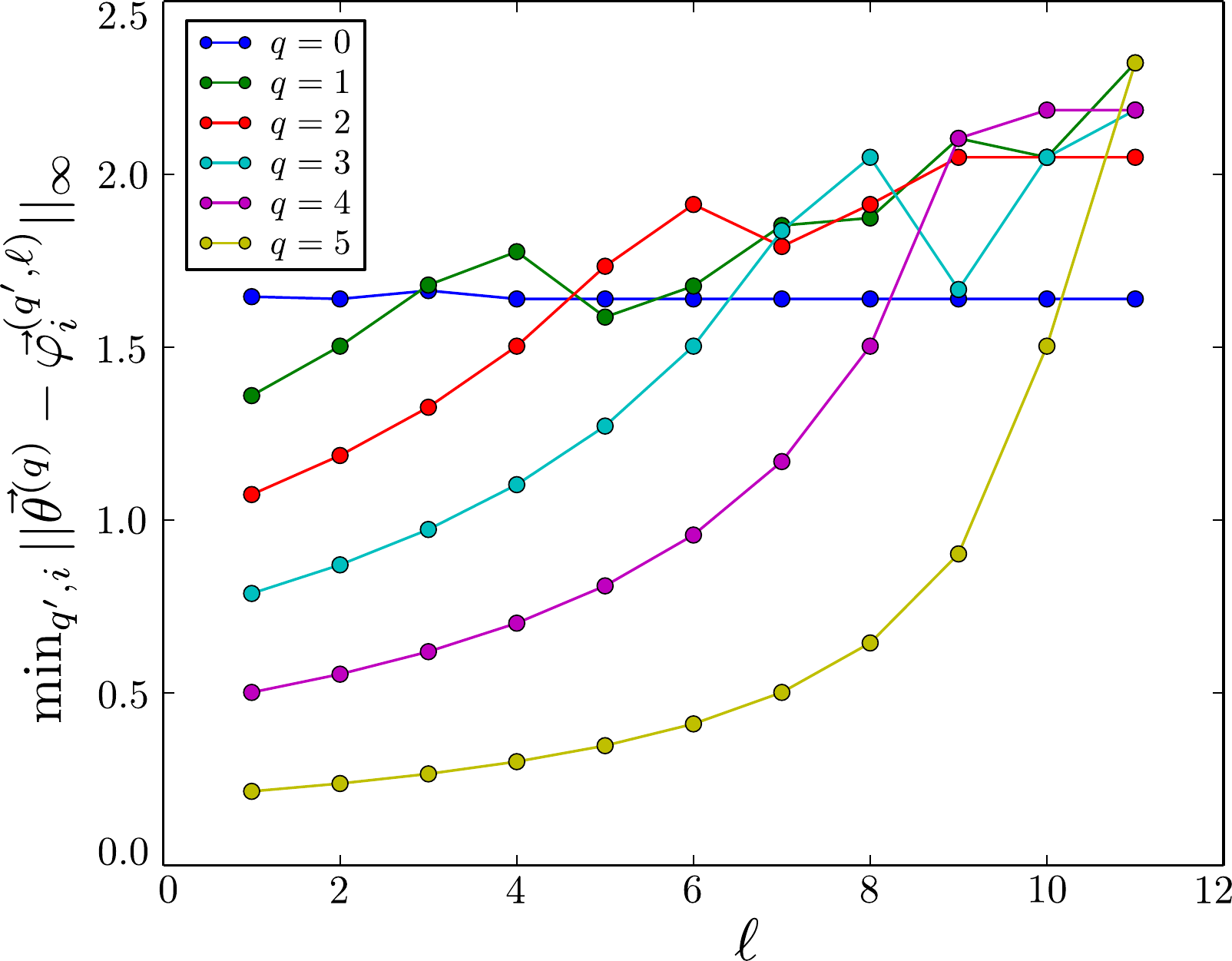}
 \caption{Distance between $\vec{\theta}^{(q)}$, the stable fixed point with winding number $q$, and the closest unstable fixed point with $\ell$ angle differences larger than $\pi/2$, 
 for a cycle of length $n=23$.}
 \label{fig:fixed_point_dist}
\end{figure}

\begin{figure}
 \centering
 \includegraphics[width=.4\textwidth]{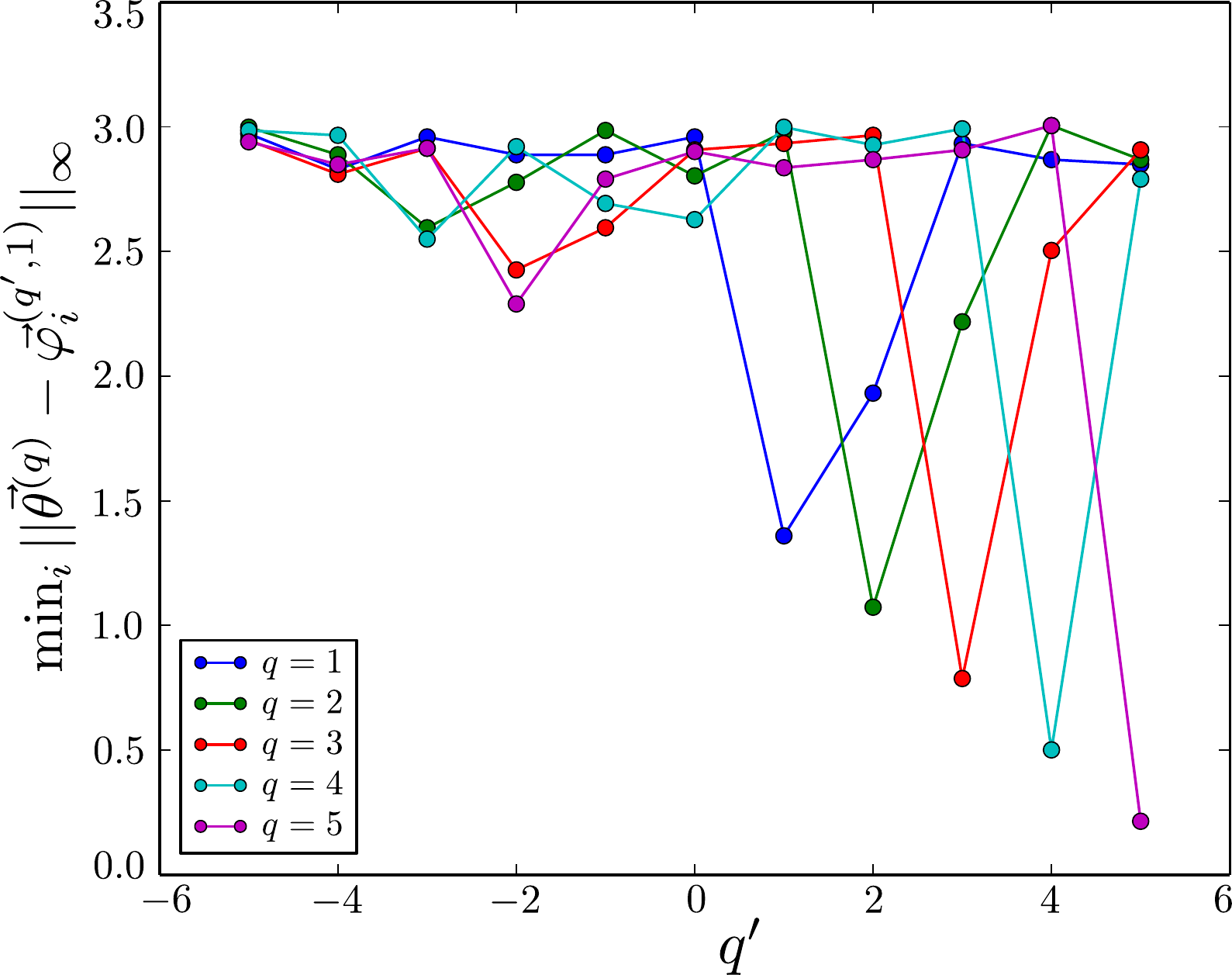}
 \caption{Distance between $\vec{\theta}^{(q)}$, the stable fixed point with winding number $q$, and the closest 1-saddle with winding number $q'$, for a cycle of length $n=23$.}
 \label{fig:fixed_point_dist2}
\end{figure}

Fig.~\ref{fig:fixed_point_dist} shows that the 1-saddles are the closest unstable fixed points to stable fixed points. 
Fig.~\ref{fig:fixed_point_dist2} shows that stable fixed points and 1-saddles are the closest if they have the same winding number.
As we remarked in Sec.~\ref{sec:analytic}, the case of $q=0$ is special because there are no unstable fixed points $\vec{\varphi}^{(0,\ell\neq0)}_i$, with winding number zero.
\end{appendices}


%

\end{document}